\documentclass[aps,amsmath,amssymb,twocolumn]{revtex4-1}

\usepackage{graphicx}
\usepackage{dcolumn}
\usepackage{bm}
\usepackage{color}

\begin{document}

\title{Super-Planckian Far-Field Radiative Heat Transfer}

\author{V. Fern\'andez-Hurtado$^{1}$}
\author{A. I. Fern\'andez-Dom\'{\i}nguez$^{1}$}
\author{J. Feist$^{1}$}
\author{F. J. Garc\'{\i}a-Vidal$^{1,2}$}
\author{J. C. Cuevas$^{1,3}$}

\affiliation{$^1$Departamento de F\'{\i}sica Te\'orica de la Materia Condensada and Condensed Matter 
Physics Center (IFIMAC), Universidad Aut\'onoma de Madrid, E-28049 Madrid, Spain}
\affiliation{$^{2}$Donostia International Physics Center (DIPC), Donostia/San Sebasti\'an 20018, Spain}
\affiliation{$^{3}$Department of Physics, University of Konstanz, D-78457 Konstanz, Germany}

\date{\today}

\begin{abstract}
We present a theoretical analysis that demonstrates that the far-field radiative heat transfer between 
objects with dimensions smaller than the thermal wavelength can overcome the Planckian limit by orders of 
magnitude. We illustrate this phenomenon with micron-sized structures that can be readily fabricated 
and tested with existing technology. Our work shows the dramatic failure of the classical theory to 
predict the far-field radiative heat transfer between micro- and nano-devices.
\end{abstract}

\maketitle

Thermal radiation is a ubiquitous physical phenomenon and its understanding is critical for 
many different technologies \cite{Modest2013}. This understanding is still largely based on 
Planck's law for black bodies \cite{Planck1914}. In particular, this law sets an upper limit 
for the radiative heat transfer (RHT) between bodies at different temperatures. However, this 
law has known limitations. One of them is its inability to describe the RHT between objects 
separated by distances smaller than the thermal wavelength ($\lambda_{\rm Th} \approx 10$ 
$\mu$m at room temperature) \cite{Joulain2005,Basu2009,Song2015}. As predicted in the 1970s
\cite{Polder1971} within the theory of fluctuational electrodynamics (FE) \cite{Rytov1989}, RHT in 
this near-field regime is dominated by evanescent waves and the Planckian limit can be far 
surpassed by bringing objects sufficiently close. The experimental verification of this prediction 
in recent years has boosted the field of thermal radiation \cite{Kittel2005,Rousseau2009,Shen2009,
Ottens2011,Kralik2012,Zwol2012a,Zwol2012b,Worbes2013,St-Gelais2014,Song2015a,Kim2015,St-Gelais2016,
Song2016,Bernardi2016} and has triggered the hope that near-field RHT can have an impact 
in different thermal nanotechnologies \cite{Basu2009,Song2015}.

As already acknowledged by Planck \cite{Planck1914}, another limitation of his law is related 
to the description of RHT between objects with dimensions smaller than 
$\lambda_{\rm Th}$. In this case, Planck’s law, which is based on ray optics, is expected to 
fail even in the far-field regime, where separations are larger than $\lambda_{\rm Th}$. Thus, one 
may wonder whether the Planckian limit can also be overcome in the far-field regime, something 
that is not possible with extended (infinite) objects \cite{Biehs2016}. It is known that the emissivity 
of a finite object can be greater than 1 at certain frequencies \cite{Bohren1998,Schuller2009}, 
but that is not enough to emit more than a black body. In fact, only a modest super-Planckian 
thermal emission has been predicted in rather academic situations \cite{Kattawar1970,Golyk2012}, 
and it has never been observed \cite{note1,Wuttke2013}. In the case of heat transfer there are neither 
theoretical proposals nor observations of super-Planckian far-field RHT. This is mainly due to 
the lack, until recently, of numerical techniques able to describe the RHT between objects of 
arbitrary size and shape that can, in turn, guide the design of appropriate experiments. This 
fundamental problem is also of great practical importance because Planck's law continues to be 
the basis for the description of far-field RHT between micro-devices \cite{Kim2001,Shi2003,Lee2015,
Lee2017,Sadat2013,Zheng2013}, which as we shall show in this work can lead to severe errors.

The goal of this work is to demonstrate that Planck's law can fail dramatically when describing the 
far-field RHT between finite objects and, in particular, that the Planckian limit can be greatly 
overcome in the far-field regime. For this purpose, we have combined state-of-the-art numerical 
simulations within FE with analytical insight provided by the general connection between the 
far-field RHT between finite objects and their radiation absorption properties. For didactic 
purposes, we first consider the case of isotropic bodies. Using a thermal discrete dipole 
approximation (TDDA) \cite{Martin2017}, we were able to show that the radiative power 
exchanged by two identical and isotropic bodies at temperatures $T_1$ and $T_2$ and separated 
by a distance much larger than both $\lambda_{\rm Th}$ and their characteristic dimensions is 
given by \cite{SM}
\begin{equation}
P = \pi A F_{12} \int^{\infty}_0 Q^2(\omega) \left[ I_{\rm BB}(\omega,T_1) -
I_{\rm BB}(\omega,T_2) \right] d\omega .
\label{eq-Q}
\end{equation}
Here, $A$ is the area of the bodies, $F_{12}$ is a geometrical view factor \cite{Modest2013}, 
$Q(\omega)$ is the frequency-dependent emissivity, which is equal to the absorption efficiency 
\cite{Bohren1998}, i.e., the normalized absorption cross-section of the bodies, and 
$I_{\rm BB}(\omega,T)$ is the Planck distribution function given by
\begin{equation}
I_{\rm BB}(\omega,T) = \frac{\omega^2}{4\pi^3 c^2} \frac{\hbar \omega}
{\exp(\hbar \omega/k_{\rm B}T) -1} ,
\end{equation}
where $\hbar$ is Planck's constant, $k_{\rm B}$ is Boltzmann's constant, and $c$ is the speed of 
light. For black bodies $Q(\omega)=1$ for all frequencies and Eq.~(\ref{eq-Q}) 
reduces to the Stefan-Boltzmann law \cite{Modest2013}: $P_{\rm BB} = \sigma A F_{12}(T^4_1-
T^4_2)$, where $\sigma = 5.67 \times 10^{-8}$ W/(m$^2$K$^4$). Equation~(\ref{eq-Q}) has the 
expected form from the expression of thermal emission of a sphere \cite{Rytov1989,Bohren1998,
Kattawar1970}, where $Q(\omega$) is independent of the direction and the polarization. However, 
this result has not been reported in the literature and its generalization to anisotropic bodies 
is non-trivial (see below). Now, since the absorption efficiency of a finite object can be larger 
than unity \cite{Bohren1998}, Eq.~(\ref{eq-Q}) suggests that super-Planckian far-field RHT might 
be possible if we find the right combination of material and object shape that leads to resonant 
absorption close to the maximum of Planck's distribution at a given temperature.  

\begin{figure*}[t]
\begin{center} \includegraphics[width=0.6\textwidth,clip]{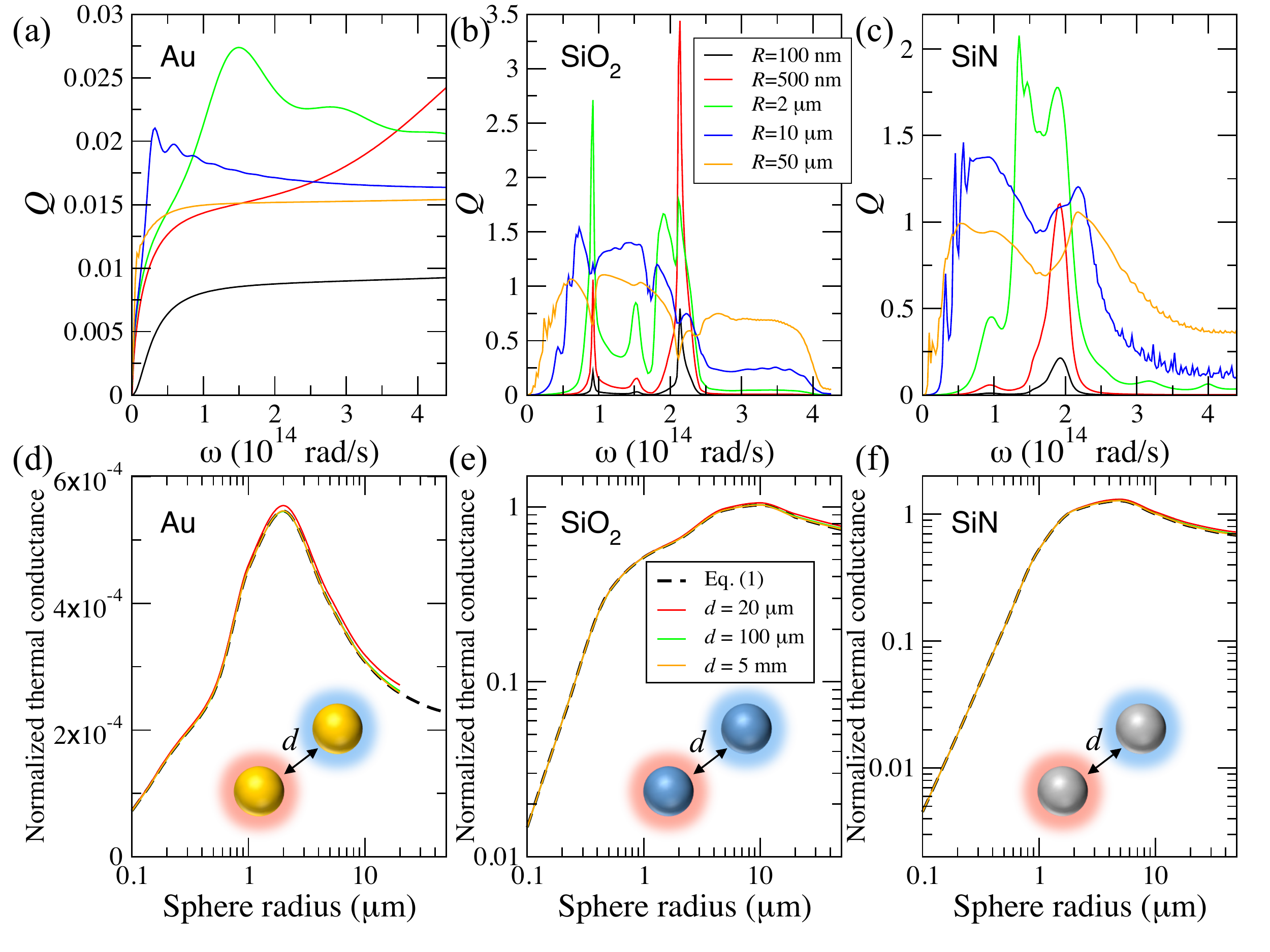} \end{center}
\caption{(Color online) (a-c) Absorption efficiency (or emissivity) of a sphere of Au (a), 
SiO$_2$ (b), and SiN (c) as a function of the frequency. The different curves correspond to 
different values of the sphere radius $R$, as indicated in the legend of panel (b). 
(d-f) Room-temperature radiative heat conductance, normalized by the blackbody results, 
for two identical spheres separated by a distance $d$ as a function of the sphere radius. 
The spheres are made of Au (d), SiO$_2$ (e), and SiN (f). The solid lines correspond to the 
exact calculations for different gaps in the far-field regime, see legend in panel (e), 
while the black dashed line corresponds to the results obtained by combining Eq.~(\ref{eq-Q}) 
with the results of panels (a-c).}
\label{fig-spheres}
\end{figure*}

This appealing idea is, however, not easy to realize in practice. We illustrate this fact in 
Fig.~\ref{fig-spheres} where we show the results for the absorption efficiency and the far-field 
RHT for spheres made of a metal (Au) and two polar dielectrics (SiO$_2$ and SiN) with radii ranging 
from 100 nm to 50 $\mu$m. The absorption efficiencies, which in this case are independent of the 
angle of incidence and polarization, were obtained with the analytical Mie theory \cite{Bohren1998}. 
The RHT is characterized here in terms of the room-temperature linear heat conductance, which is 
normalized in Fig.~\ref{fig-spheres} by the corresponding blackbody result: $G_{\rm BB} = 4
\sigma A F_{12} T^3$. We present the results computed with Eq.~(\ref{eq-Q}) and the absorption 
efficiencies as well as numerical results for three gaps in the far-field regime (20 $\mu$m, 
100 $\mu$m, and 5 mm) that were obtained with the code SCUFF-EM \cite{Rodriguez2013,Reid2015}. 
This code implements a fluctuating-surface-current formulation of the RHT problem and provides 
numerically exact results within the framework of fluctuational electrodynamics. 

The results of Fig.~\ref{fig-spheres}(d-f) show that the far-field RHT between spheres 
does not overcome the Planckian limit and only in the case of the polar dielectrics it becomes 
comparable to the blackbody result when the sphere radius is of the order of $\lambda_{\rm Th}$. For 
small radii, smaller than the corresponding skin depth, the normalized conductance increases linearly 
with the radius, i.e., the conductance is proportional to the sphere volume because the whole particle 
contributes to the RHT. In the opposite limit, when the radius is much larger than $\lambda_{\rm Th}$ the 
conductance tends to the result for two parallel plates. It is worth stressing that the numerical 
results obtained for various gaps nicely confirm the validity of Eq.~(\ref{eq-Q}).

\begin{figure}[t]
\begin{center} \includegraphics[width=\columnwidth,clip]{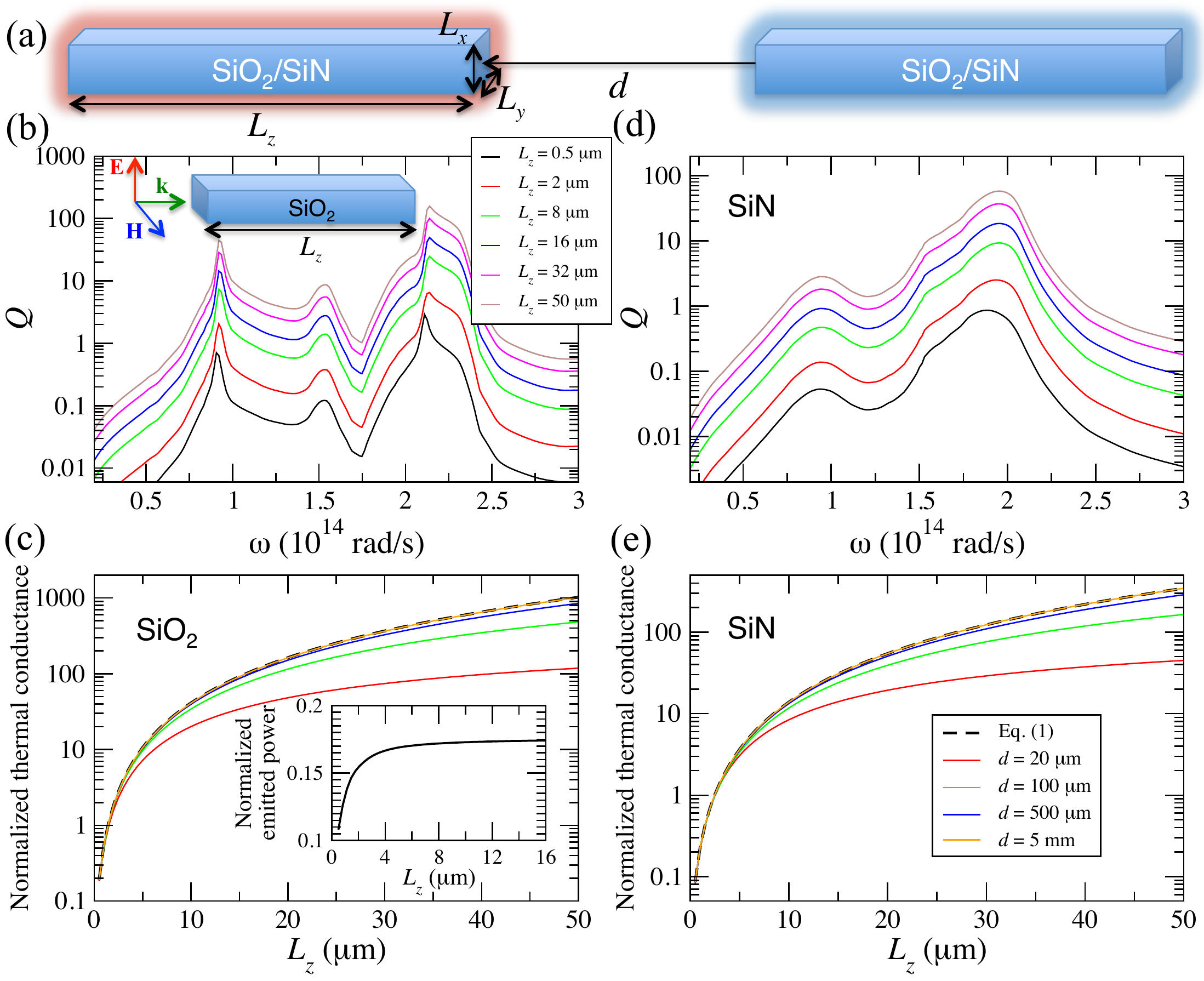} \end{center}
\caption{(Color online) (a) Two identical parallelepipeds made of SiO$_2$ or SiN with dimensions 
$L_x = L_y = 0.5$ $\mu$m and varying $L_z$ and separated by a gap $d$. (b) Absorption efficiency 
as a function of the frequency for a linearly polarized plane wave impinging with normal incidence,
see inset, in a SiO$_2$ parallelepiped with $L_x = L_y = 0.5$ $\mu$m and various values of $L_z$,
see legend. (c) Room-temperature radiative heat conductance, normalized by the blackbody 
results, for the system of panel (a) as a function of $L_z$. The solid lines correspond to the 
exact calculations for different gaps, see legend in panel (e), while the black dashed line is the 
result obtained with Eq.~(\ref{eq-Q}). The inset shows the total power emitted by a single SiO$_2$ 
parallelepiped with dimensions $L_x = L_y = 0.5$  $\mu$m and varying $L_z$ at room temperature. 
The result is normalized by the power emitted by a black body. (d-e) The same as in panels (b) and 
(c), but for SiN.}
\label{fig-wires}
\end{figure}

The results for spheres, and for other geometries like cubes \cite{SM}, show that although the 
absorption efficiency (or emissivity) can be larger than 1 for some frequencies, this does not 
imply a super-Planckian far-field RHT. Then, what is the strategy to overcome the Planckian limit? 
The answer is indeed provided by Eq.~(\ref{eq-Q}), which suggests that far-field RHT can be enhanced 
by increasing the absorption cross section, while maintaining the geometrical one. This idea is 
illustrated in Fig.~\ref{fig-wires} where we show the far-field RHT between two parallelepipeds 
of SiO$_2$ and SiN as well as the relevant emissivities, which are those related to the direction 
joining the parallelepipeds. Here, we start with a cube of side 0.5 $\mu$m (much smaller than 
$\lambda_{\rm Th}$) and we form an elongated parallelepiped by progressively changing one of the dimensions, 
$L_z$, while keeping constant the other two, $L_x$, and $L_y$, see Fig.~\ref{fig-wires}(a). This 
way we keep the relevant geometrical cross section constant, while the absorption cross section 
increases linearly with $L_z$ as long as it is small compared to the penetration depth of the radiation, 
which is hundreds of microns. Our calculations of these emissivities, which were performed with
the TDDA approach \cite{Martin2017}, indeed confirm that they can become much larger than 1 in a 
broad frequency range, as we show in Fig.~\ref{fig-wires}(b,d). This fact naturally leads to far-field 
conductance values that overcome the Planckian limit by several orders of magnitude when $L_z$ becomes 
of the order of 50 $\mu$m, see Fig.~\ref{fig-wires}(c,e). Notice that the numerical results obtained
with TDDA \cite{Martin2017} for several gaps in the far-field regime coincide with those obtained via 
Eq.~(\ref{eq-Q}) in the limit of sufficiently large gaps. In any case, the Planckian limit is greatly 
overcome for all gaps in the far-field regime. Let us also stress that these parallelepipeds are not 
super-Planckian thermal emitters, as we show in Fig.~\ref{fig-wires}(c). The super-Planckian RHT found 
here is possible due to the highly directional emission of these systems. 

\begin{figure}[t]
\begin{center} \includegraphics[width=\columnwidth,clip]{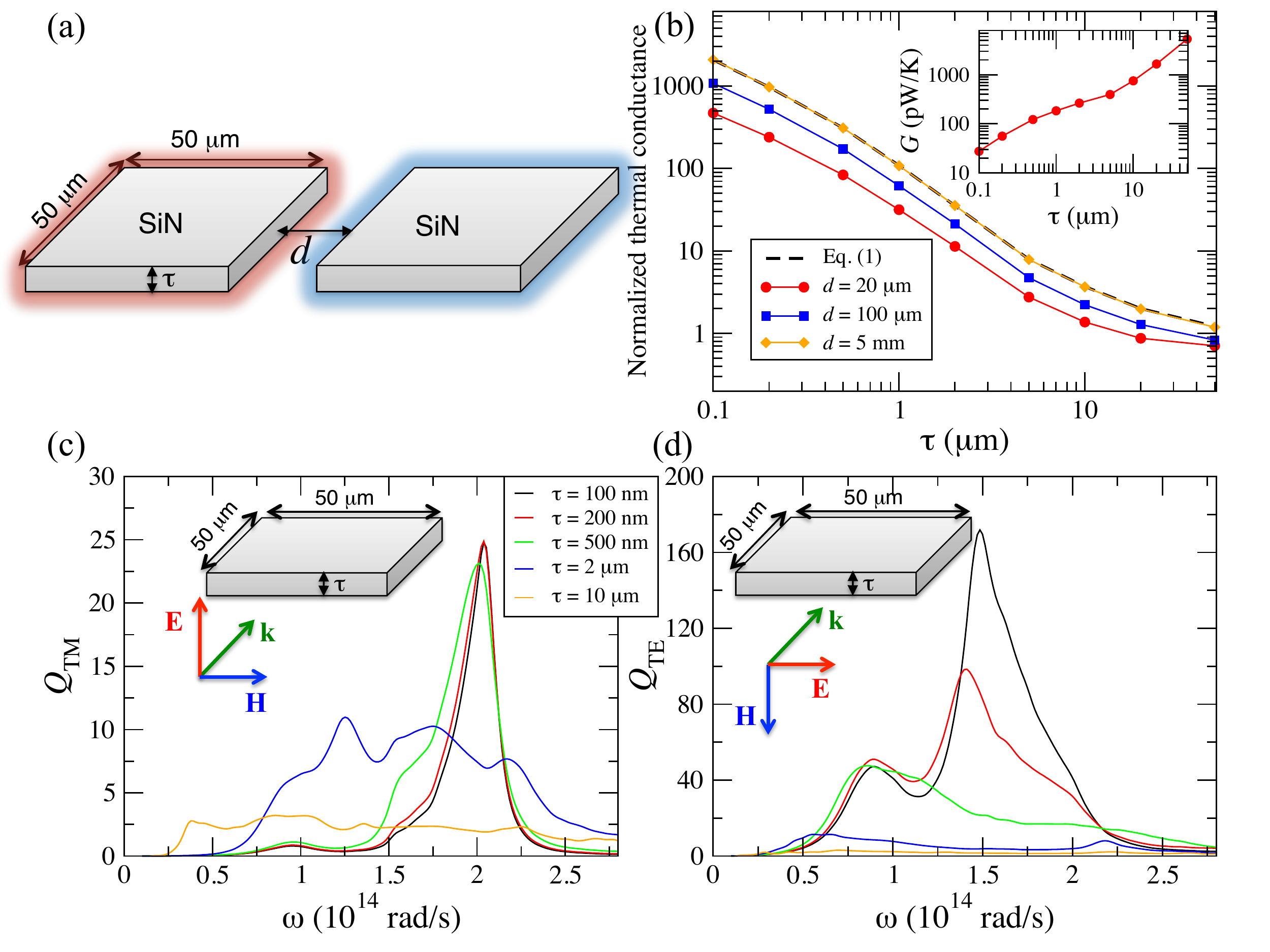} \end{center}
\caption{(Color online) (a) SiN pads with lateral dimensions of $50 \times 50$ $\mu$m$^2$, 
a thickness $\tau$, and separated by a gap $d$. (b) Room-temperature radiative heat conductance, 
normalized by the blackbody results, for the system of panel (a) as a function of the pad thickness. 
The solid lines correspond to the exact calculations for three gaps (see legend) and the black dashed
line is the result obtained by combining Eq.~(\ref{eq-Q}) with the results of panels (c) and (d). The 
inset shows the results for $d = 20$ $\mu$m without normalization. (c) Absorption efficiency as a 
function of the frequency for a plane wave impinging with normal incidence and transverse magnetic 
(TM) polarization in a SiN pad with lateral dimensions of $50 \times 50$ $\mu$m$^2$ and various 
thicknesses, as indicated in the legend. The inset describes this polarization. (d) The same as in 
panel (c), but for a transverse electric (TE) polarization.}
\label{fig-pads}
\end{figure}

These results clearly demonstrate the severe failure of the classical theory of thermal radiation, 
but the absolute conductance values for these tiny parallelepipeds make their measurement challenging. 
To illustrate super-Planckian heat transfer in a system that can be tested with existing technology, 
we focus now on the analysis of the RHT between two identical SiN pads with fixed lateral dimensions 
of $50 \times 50$ $\mu$m$^2$, see Fig.~\ref{fig-pads}(a), which are larger than $\lambda_{\rm Th}$ 
at room temperature, and we vary their thickness, $\tau$, from values much smaller than $\lambda_{\rm Th}$ 
to values larger than this wavelength. This challenging system for the theory is inspired by the 
suspended-pad micro-devices that are widely employed for measuring thermo-physical properties of 
low-dimensional nanostructures (nanotubes, nanowires or nanoribbons) \cite{Kim2001,Shi2003,Lee2015,Lee2017}. 
These devices consist of two adjacent SiN membranes suspended by long beams. Each membrane features 
a platinum resistance heater/thermometer that is normally used to measure the heat conduction 
through a sample that bridges the gap between the membranes, but they can also be used to measure 
the RHT across the gap. In recent years, these micro-devices have reached sensitivities of $\sim$1 
pW/K and below \cite{Sadat2013,Zheng2013}.

\begin{figure}[!t]
\begin{center} \includegraphics[width=\columnwidth,clip]{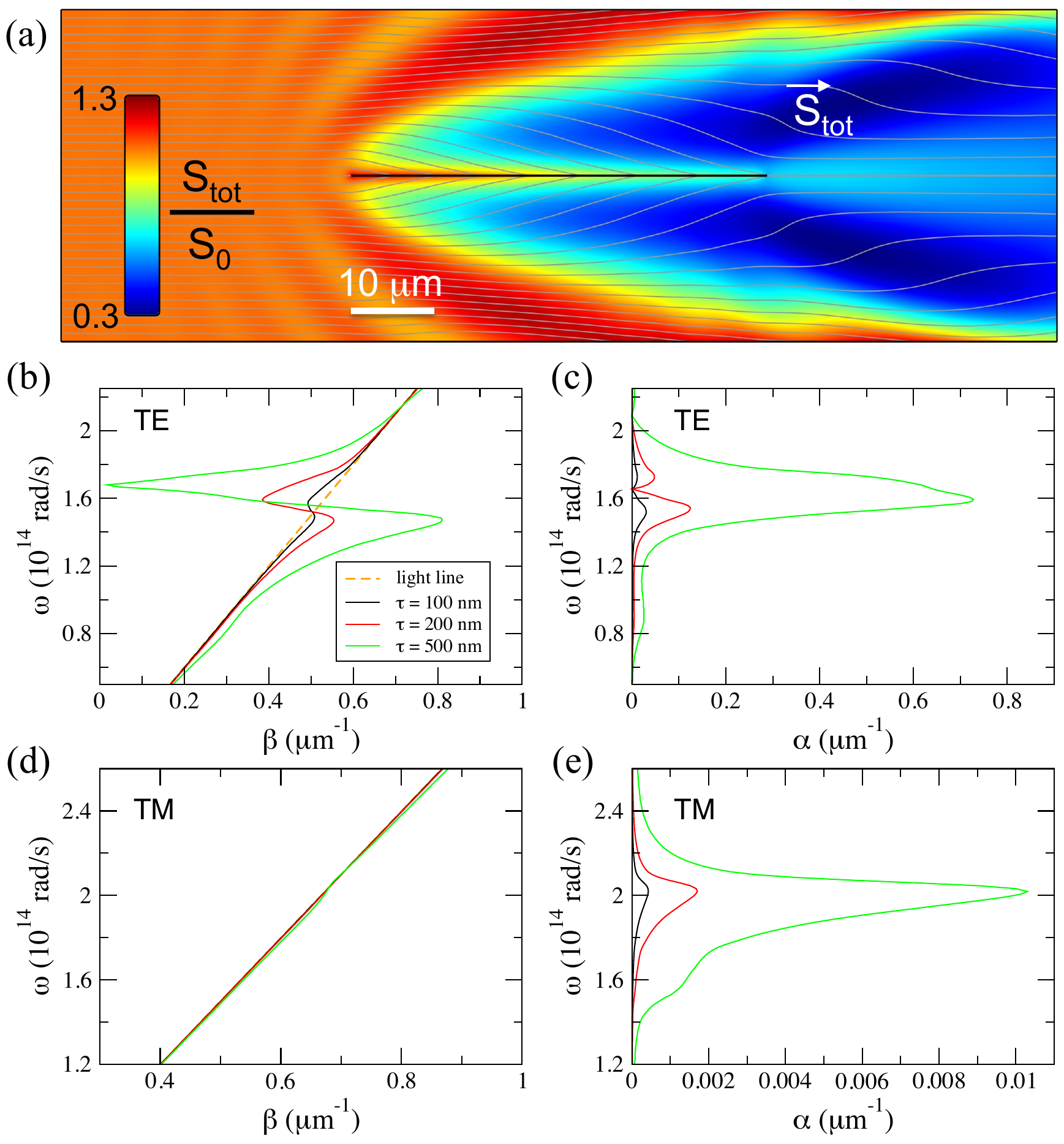} \end{center}
\caption{(Color online) (a) Magnitude of the total Poynting vector $S_{\rm tot}$, normalized by 
the incident value $S_0$, in a simulation of a TE-polarized plane wave impinging from the left in 
a 100 nm-thick SiN pad (in black) for a frequency of $1.5 \times 10^{14}$ rad/s, which corresponds 
to the maximum of the absorption efficiency. The pad length is 50 $\mu$m, while it is assumed to be 
infinite in the transverse direction. A streamline representation of the Poynting vector is shown 
in grey solid lines. (b-c) Frequency versus propagation constant $\beta$ (b) and attenuation constant
$\alpha$ (c) for the TE modes of an infinite SiN slab for different values of its thickness $\tau$.
For these values of $\tau$, the SiN slab supports a single mode for every frequency. The dashed 
line in panel (b) is the light line in vacuum, $\omega = \beta c$, and it separates the modes into 
guided ones on the right hand side of this line and leaky ones on the left. (d-e) The same as in 
panels (b) and (c) for TM modes.}
\label{fig-modes}
\end{figure}

Equation~(\ref{eq-Q}) still applies for parallelepipeds if we replace $Q^2(\omega)$ by $[Q^2_{\rm TM}
(\omega) + Q^2_{\rm TE}(\omega)]/2$ \cite{SM}, where $Q_{\rm TM,TE}(\omega)$ are the absorption efficiencies 
for a plane wave with normal incidence and transverse magnetic (TM) or transverse electric (TE) polarization, 
see insets of Fig.~\ref{fig-pads}(c,d). The results obtained with COMSOL for these efficiencies for thicknesses 
varying from 100 nm to 50 $\mu$m are displayed in Fig.~\ref{fig-pads}(c,d) and they show that when 
$\tau \ll \lambda_{\rm Th}$ they reach values of up to several hundreds, especially for TE polarization. 
Using these results together with the modified Eq.~(\ref{eq-Q}), we have computed the radiative thermal 
conductance for different values of $\tau$, see Fig.~\ref{fig-pads}(b). The ratio with the blackbody result 
increases monotonically as $\tau$ decreases, becoming clearly larger than 1 when $\tau < \lambda_{\rm Th}$, 
and reaching values as large as 2000 for a thickness of 100 nm. The occurrence of super-Planckian far-field 
RHT is confirmed by our SCUFF-EM results for several gap, as shown in Fig.~\ref{fig-pads}(b). These results 
show that, irrespective of the gap, the super-Planckian RHT takes place when $\tau < \lambda_{\rm Th}$ 
and that it should be readily observed in suspended-pad micro-devices, as we illustrate in the inset
of Fig.~\ref{fig-pads}(b). They also show that Eq.~(\ref{eq-Q}), with the modification explained above,
provides the asymptotic result for gaps much larger than the dimensions of the objects, which is when 
the largest enhancements over the blackbody theory occur. It is worth stressing that, following the same 
strategy as for the parallelepipeds of Fig.~\ref{fig-wires}, this super-Planckian RHT can be further enhanced 
by increasing the depth of the pads \cite{SM}. Let us also stress that, as in the case of the parallelepipeds,
these pads are not super-Planckian emitters, as we show in Ref.~[\onlinecite{SM}].

The remarkable absorption efficiency and the super-Planckian behavior in these pads are due to the 
fact that they behave as (lossy) dielectric waveguides that absorb the radiation via the excitation 
of guided modes. This excitation results first in a confinement of the radiation field, as we 
illustrate in the COMSOL simulation shown in Fig.~\ref{fig-modes}(a). Then, due to the low-impedance 
mismatch, the incident radiation couples efficiently into guiding modes and is eventually absorbed. 
This absorption can be understood with an analysis of the electromagnetic modes of a planar waveguide, 
where an infinite SiN slab (core) of thickness $\tau$ is surrounded by air (cladding). Using dielectric 
waveguide theory \cite{Marcuse1991,SM}, we have computed for each polarization (TM and TE) the real 
(propagation constant) and imaginary part (attenuation constant) of the wave vector component along 
the slab of these modes. The results are shown in Fig.~\ref{fig-modes}(b-e) for several thicknesses 
in the regime where there is a single mode. The attenuation constant, which determines the power 
absorption, strongly depends on the polarization and exhibits much larger values for the TE case, 
which explains the larger efficiency for this polarization, see Fig.~\ref{fig-pads}(d). Moreover, the 
frequency dependence of the attenuation constant reproduces the corresponding dependence of the 
absorption efficiencies for both polarizations in the limit of thin pads. Interestingly, there is 
a frequency range in which the slab behaves as a hollow dielectric waveguide \cite{Marcuse1991}, 
with the real part of the core dielectric function being smaller than that of the cladding, and the 
propagation for TE polarization occurs via leaky modes \cite{Marcuse1991}, see Fig.~\ref{fig-modes}(b,c).

So in summary, our work illustrates the need to revisit the far-field RHT between micro- and nano-systems 
in the light of fluctuational electrodynamics. In particular, we have shown the dramatic failure of 
Planck's law in micron-sized devices of great importance for the field of thermal transport 
\cite{Kim2001,Shi2003,Lee2015,Lee2017,Sadat2013,Zheng2013}, whose sensitivity is ultimately limited by 
thermal radiation. Our work is also important for the study of thermalization of small objects
\cite{Wuttke2013} with implications, e.g., in cavity optomechanics experiments \cite{Chang2010}
or in the study of interstellar dust in astrophysics \cite{Draine2011}. Our work also raises the 
question of the ultimate limit of the super-Planckian far-field RHT in low-dimensional systems 
such as nanowires, nanoribbons or 2D materials.

We thank A. Garc\'{\i}a-Mart\'{\i}n, E. Moreno, and Y. Shi for fruitful discussions. We acknowledge 
funding from the Spanish MINECO (FIS2015-64951-R, MAT2014-53432-C5-5-R, FIS2014-53488-P), 
the Comunidad de Madrid (S2013/MIT-2740), the European Union Seventh Framework Programme 
(FP7-PEOPLE-2013-CIG-630996, FP7-PEOPLE-2013-CIG-618229), and the European Research Council 
(ERC-2011-AdG-290981 and ERC-2016-STG-714870). V.F.-H.\ acknowledges support from “la Caixa” 
Foundation and J.C.C.\ thanks the DFG and SFB767 for sponsoring his stay at the University 
of Konstanz as Mercator Fellow.


\end{document}